\documentclass{article}
\usepackage{PRIMEarxiv}
\usepackage[utf8]{inputenc} 
\usepackage[T1]{fontenc}    
\usepackage{hyperref}       
\usepackage{url}            
\usepackage{booktabs}       
\usepackage{amsfonts}       
\usepackage{nicefrac}       
\usepackage{microtype}      
\usepackage{wrapfig}
\usepackage{fancyhdr}       
\usepackage{graphicx}       
\usepackage{float}
\usepackage{caption}
\usepackage{placeins}
\usepackage{csvsimple}
\usepackage{booktabs}
\usepackage{pgfplotstable}
\usepackage[round]{natbib}
\usepackage{enumitem}

\pgfplotstableset{col sep=comma, string type}

\graphicspath{{media/}}     
\usepackage{amsmath}
\usepackage{amssymb}
\usepackage{verbatim}

\usepackage{float}
\restylefloat{figure}
\restylefloat{table}
\title{Noise-proofing Universal Portfolio Shrinkage
}

\author{
  Paul Ruelloux$^{1,2}$, Christian Bongiorno$^{1}$, Damien Challet$^{1}$, \\[0.3em]
  $^{1}$Université Paris-Saclay, Laboratoire de Mathématiques et Informatique pour la Complexité et les Systèmes (MICS)\\
  $^{2}$Barclays Bank Ireland PLC\\
  [0.3em]
  \texttt{\{name\}.\{surname\}@centralesupelec.fr}
}

\begin{document}

\maketitle

\begin{abstract}  
We enhance the Universal Portfolio Shrinkage Approximator (UPSA) of \cite{NBERw32004} by making it more robust with respect to estimation noise and covariate shift. UPSA optimizes the realized Sharpe ratio using a relatively small calibration window, leveraging ridge penalties and cross-validation to yield better portfolios. Yet, it still suffers from the staggering amount of noise in financial data. We propose two methods to make UPSA more robust and improve its efficiency:  time-averaging of the optimal penalty weights and using the Average Oracle correlation eigenvalues to make covariance matrices less noisy and more robust to covariate shift. Combining these two long-term averages outperforms UPSA by a large margin in most specifications.
\end{abstract}

\section{Introduction}
Markowitz portfolio theory \citep{Markowitz} prescribes efficient portfolios that achieve maximum risk-adjusted performance, but the quality of such portfolios depends critically on the accuracy of the estimated covariance matrix \citep{michaud1989markowitz}. In practice, estimation noise severely undermines performance, especially when the number of assets $n$ is of the same order as the sample size $\Delta t$, i.e., in the high-dimensional case: sample covariance matrix estimates become very noisy and require filtering. Current state-of-the-art shrinkage methods address this problem by transforming the in-sample (IS) eigenvalues to minimize the expected mean-squared error (MSE) between the sample and population covariance matrices, thereby indirectly improving the out-of-sample (OOS) portfolio performance \citep{bun2017cleaning, ledoit2009eigenvectorslargesamplecovariance}.

The Universal Portfolio Shrinkage Approximator (UPSA), proposed by \cite{NBERw32004}, is a flexible nonlinear spectral shrinkage method that aims to maximize the expected OOS Sharpe ratio. Its robustness comes from Cross-Validation (CV) and weighted set of portfolios associated to ridge penalization factors. While improves upon simpler shrinkage techniques,  it still suffers from sample noise: the optimal ridge portfolio weights tend to be concentrated on a single one, which itself varies much as a function of time. This comes from the fact that CV still leaves too much noise in high-dimensional settings and cannot account for covariate shift \citep{Cawley2010,Arlot2010}. As a consequence, the performance of UPSA substantially depends on the ridge penalty grid and can be improved by further filtering. 

We propose two simple remedies that boost UPSA performance by reducing the influence of sample noise. First, we replace sample-specific optimal weights of ridge-penalized portfolios with their time averages (computed over expanding time windows). Second, we apply a pre-filtering of correlation eigenvalues known as the Average Oracle (AO) method \citep{bongiorno2023filtering}, which also replaces local filtering with time averages of optimal eigenvalues; this method is known to over-perform CV \citep{bongiorno2023filtering,AO2}. Combining both improvements brings better performance than either UPSA or AO alone when applied on monthly factor returns. 

This letter is organized as follows. Section 2 introduces the rationale and construction of UPSA and discusses its limitations and introduces the proposed remedies. Section 3 reports empirical results across different configurations. Section 4 concludes and proposes to interpret UPSA as first-order improvement of Average Oracle.

\section{Estimators}
\subsection{Universal Portfolio Shrinkage Approximator (UPSA)}
The objective of a shrinkage approach is to obtain an improved estimate of the unknown population covariance matrix $\boldsymbol{\Sigma}$ from a finite-sample covariance estimator 
$\hat{\boldsymbol{\Sigma}} = \frac{1}{\Delta t}(\boldsymbol{X} - \hat{\boldsymbol{\mu}})(\boldsymbol{X} - \hat{\boldsymbol{\mu}})^\top$, 
where $X$ denotes the $n \times \Delta t$ matrix of observations with true mean $\boldsymbol{\mu}$ and covariance $\boldsymbol{\Sigma}$. 
The method proceeds from the spectral decomposition of $\hat{\boldsymbol{\Sigma}}$, expressed as 
$\hat{\boldsymbol{\Sigma}} = \hat{\boldsymbol{V}} \hat {\boldsymbol{\Lambda}} \hat{\boldsymbol{V}}^\top$, 
where $\hat{\boldsymbol{V}}$ is the orthogonal matrix of eigenvectors and $\hat {\boldsymbol{\Lambda}}$ is the diagonal matrix of sample eigenvalues $\boldsymbol{\lambda}=\{\lambda_k \}_{k=1}^n$. 
Shrinkage consists in replacing the empirical eigenvalues $\hat \lambda_k$ with shrunk (adjusted) values $\hat f( \hat \lambda_k)$, 
where $\hat f$ is a shrinkage function that optimizes a well-chosen cost function: while the so-called Non-Linear Shrinkage (NLS) estimators minimize the estimation error of $\hat {\boldsymbol{\Sigma}}$ \citep{ledoit2009eigenvectorslargesamplecovariance}, UPSA aims at maximizing the realized Sharpe ratio. Both methods use data from the calibration window only. As a consequence, the shrinkage function $f$ itself is subject to noise and is denoted by $\hat f$.

The resulting covariance estimator takes the general form
\begin{equation}\label{eq:rie}
    \boldsymbol{\hat{\Xi}} = \boldsymbol{\hat{V}} \,\textrm{Diag }(\hat f(\hat{\boldsymbol{ \lambda }})) \boldsymbol{\hat{V}^\top},
\end{equation}
where we used a slight abuse of notation: $\hat{f}(\hat{\boldsymbol{\lambda}})$ denotes the vector of shrunk eigenvalues  $\hat f(\hat \lambda_i)$.

In the general formulation, a shrinkage function $f : \mathcal{I} \subset \mathbb{R}_+ \rightarrow \mathbb{R}_+$ can be expressed as a function of a positive finite measure $\nu$ on $\mathbb{R}_+$ as
\begin{equation}
\label{eq:shrinkage_integral_form}
    \forall \lambda \in \mathbb{R}_+, \quad  
    f(\lambda) = \int_{0}^{\infty} \frac{1}{z + \lambda}\, d\nu(z).
\end{equation}
Thus, finding the optimal $\hat f$  given the calibration data and a choice of loss function is equivalent to finding the optimal $\hat \nu$.
\cite{NBERw32004} replace the integral by a weighted combination of ridge-penalized terms $\zeta_{i k}:=(\lambda_k+z_i)^{-1}$
\begin{equation}\label{eq:ridgeshrink}
    f(\lambda) \approx  f(\lambda_k \mid \mathbf{z},\boldsymbol{ \alpha}) 
   =\sum_{i=1}^{\ell} \alpha_i \zeta_{i k} =\sum_{i=1}^{\ell} \frac{\alpha_i}{z_i + \lambda_k}, \quad \text{with}\quad   k = 1, \dots, n,
\end{equation}
where  $z_i$ are the ridge penalties associated with each weight $\alpha_i$.  Given a set of ridge penalties $\mathbf{z}=\{z_i\}_{i=1}^\ell$, the weights $\boldsymbol{\alpha}=\{\alpha_i\}_{i=1}^\ell$  are chosen so as to optimize the loss function.

Let us focus on the Sharpe ratio-maximizing procedure of UPSA: the mixture weights $\boldsymbol{\alpha}$ are estimated through a leave-one-out procedure that separates IS estimation from OOS evaluation. For each fold and each ridge level $z_i$, the method applies the eigenvalue shrinkage defined by $\zeta_{ik }:=(\lambda_k+z_i)^{-1}$ to the IS covariance estimator $\hat{\boldsymbol{\Xi}}$ and, using the resulting filtered precision matrix together with IS means, computes the IS maximum–Sharpe-ratio portfolio.  This portfolio is then evaluated on the held-out data to obtain its OOS return. Averaging these basis returns over all the folds yields the OOS mean vector and covariance matrix of basis returns, denoted $\hat{\boldsymbol{m}}(\mathbf{z}) \in \mathbb{R}^{\ell}$and $\hat{\boldsymbol{S}}(\mathbf{z}) \in \mathbb{R}^{\ell \times \ell}$. The weights $\boldsymbol{\alpha}$ are then chosen to maximize a concave quadratic objective aligned with the Sharpe criterion, yielding pseudo OOS optimal ridge weights:
\begin{equation}
\label{eq:upsa_qp_simplified}
    \hat{\boldsymbol{\alpha}}_{\mathrm{UPSA}}
    =
    \arg\max_{\boldsymbol{\alpha}\ge 0,\;\mathbf{1}^\top\boldsymbol{\alpha}=1}
    \;\;
    \boldsymbol{\alpha}^\top \hat{\boldsymbol{m}}(\mathbf{z})
    -
    \frac{1}{2}\,
    \boldsymbol{\alpha}^\top \hat{\boldsymbol{S}}(\mathbf{z})\,   \boldsymbol{\alpha}.
\end{equation}

Because all moments entering this program are computed on held-out data via the leave-one-out construction, the procedure aims to improve OOS performance. 

Finally, the UPSA weights $\hat{\boldsymbol{\alpha}}_{\mathrm{UPSA}}$ are be plugged in Eq.~\eqref{eq:ridgeshrink} to recover the filtered covariance as in Eq.~\eqref{eq:rie}, 
\begin{equation}
    \hat f(\lambda\mid\mathbf{z},\hat{\boldsymbol{\alpha}}_{\mathrm{UPSA}})
    =
    \sum_{i=1}^{\ell}\frac{\hat{\alpha}_{i,\mathrm{UPSA}}}{z_i+\lambda},
    \qquad\mathrm{thus,}\qquad
    \hat{\boldsymbol{\Xi}}_{\mathrm{UPSA}}
    =
    \hat{\boldsymbol{V}}\,\textrm{Diag } ( \hat f(\hat{\boldsymbol{\Lambda}} \mid\mathbf{z},\hat{\boldsymbol{\alpha}}_{\mathrm{UPSA}})\,)\hat{\boldsymbol{V}}^\top .
\end{equation}

Keeping all the hats was notation-wise heavy but rewarding as it made the sources of noise explicit.  In fact, we have neglected the fact that the choice of the grid $\mathbf{z}$ may also depend on the calibration data.

\subsection{Noise-proofing UPSA}\label{sec:noise}
\begin{figure}
    \begin{center} 
        \includegraphics[width=0.75\textwidth]{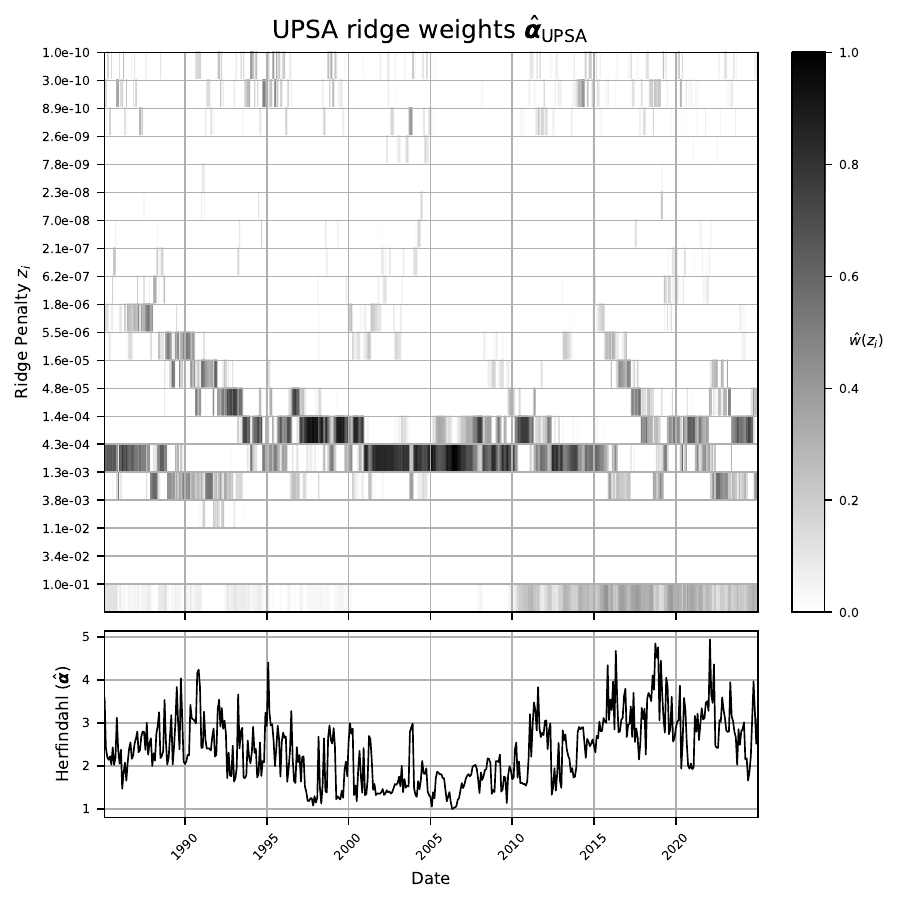} 
        \caption{Top : Heatmap of UPSA ridge weights in time computed on JKP US factors' monthly returns, with $n=153$ factors, with a ridge penalty grid of $\ell = 20$ log-spaced points in $[10^{-10},10^{-1}]$. Bottom:  Herfindahl Index (HI) vs time.
        \\}
        \label{fig:scattering_ridge} 
    \end{center}
\end{figure}

While UPSA is mathematically sound, it can be further improved. A strong (self-imposed) constraint in UPSA is the use of a relatively short calibration window. This choice mechanically leads to quite noisy outcomes, especially when computing optimal quantities, even with a leave-one-out approach. Figure \ref{fig:scattering_ridge} displays the UPSA weights $\hat{\boldsymbol{\alpha}}_{\mathrm{UPSA}}$  as a function of time, computed on the same dataset as \cite{NBERw32004}, described below in section \ref{sec:sec3}. We also report the Herfindal Index, a measure of weight concentration, defined here as $1/\sum_{i=1}^\ell |\alpha_i|^2$; it equals 1 for perfectly concentrated vectors, and $\ell$ for uniform weights \citep{Sleuwaegen1986TheHI}. The average absolute change of weights between two calibrations is about 0.45. In short, the weights are highly concentrated on values that change very often.

\subsubsection*{Averaged UPSA Ridge Weights (AvgUPSA)}
The most basic way to mitigate noisy weights is to replace them with a time average (with expanding window): at time $t$, Eq. \eqref{eq:upsa_qp_simplified} ,  $\hat{\boldsymbol{\alpha}}_{\mathrm{UPSA},t}$ becomes
\begin{equation}
    \hat{\boldsymbol{\alpha}}_{\mathrm{AvgUPSA}, \, t} = \frac{1}{t-t_0} \sum_{t'=t_0}^{t}\hat{\boldsymbol{\alpha}}_{\mathrm{UPSA}, \, t'} \, .
\end{equation}

This solves the noise in the $t\gg t_0$ limit but is bound to lack reactivity.

\subsubsection*{Average Oracle (AO)}

A major cause of noise in the UPSA weights is the estimation of the  covariance matrices. While LOO does filter some noise out, it is far less efficient than other cross-validation techniques in this context \citep{lamrani2025optimal}. In addition, if at all possible, one should account for the typical way covariance matrices change between calibration window and the test window due to covariate shift. This is the idea behind the Average Oracle, which replaces the sample eigenvalues by rank-wise averages of past "oracle" eigenvalues: it uses the data from past test windows to compute optimal eigenvalues and thus accounts for some covariate shift \citep{bongiorno2023filtering}.   
In order to calibrate the oracle eigenvalues, one considers many consecutive calibration/test sub-windows of a long IS period. For one such pair indexed by $b$, $(I_{\mathrm{cal},b}, I_{\mathrm{test},b})$ of corresponding sample covariance matrices $(\hat{ \boldsymbol{\Sigma}}_{\mathrm{cal},b}, \hat{ \boldsymbol{\Sigma}}_{\mathrm{test},b})$, one computes the vector of Oracle eigenvalues:
\begin{equation}
    \hat{\boldsymbol{\lambda}}_\mathrm{O,b} =
    \mathrm{Diag} \left( \hat{ \boldsymbol{V}}_{\mathrm{cal},b}^\top \hat{ \boldsymbol{\Sigma}}_{\mathrm{test},b} \hat{ \boldsymbol{V}}_{\mathrm{cal},b}\right),
\end{equation}
where $\hat{V}_{\mathrm{cal}}$ is the eigenvector matrix of $\hat \Sigma_{\mathrm{cal}}$.  This choice corresponds to the optimal rotationally invariant shrinkage of $\hat \Sigma_{\mathrm{cal}}$ for predicting $\hat \Sigma_{\mathrm{test}}$ \citep{ledoit2009eigenvectorslargesamplecovariance}. \footnote{This result stems from minimizing the Frobenius norm $||\boldsymbol{\Sigma}_{\mathrm{next}} - \boldsymbol{\Xi}(\boldsymbol{\Sigma}_{\mathrm{next}}) ||_{\mathrm{F}} $.}
The Average Oracle eigenvalues are then obtained as rank-wise averages of calibrated oracle eigenvalues: having computed $B\gg 1$ oracle eigenvalues:
\begin{equation}
    \hat{ \boldsymbol{\lambda}}_\mathrm{AO} = \frac{1}{B}\sum_{b=1}^B \hat{ \boldsymbol{\lambda}}_\mathrm{O,b}.
\end{equation}
To avoid look-ahead bias, only oracles computed in the strict past of each evaluation date are included in the averaging. The resulting shrinkage operator is thus: 
\begin{equation}
    \hat{ \boldsymbol{\Xi}}_{\mathrm{AO}}(\hat{ \boldsymbol{\Sigma}}) = \hat{ \boldsymbol{V}} \, \textrm{Diag }( \hat{ \boldsymbol{\lambda}}_{\mathrm{AO}}) \, \hat{ \boldsymbol{V}}^\top.
\end{equation}

 AO has been shown to be an effective zeroth-order shrinkage estimator, outperforming state-of-the-art nonlinear shrinkage estimators in many time-varying settings, including large portfolios of equity data \citep{bongiorno2023filtering,AO2}. In addition, by removing entirely the noise of eigenvalues, it leads to much less time-varying portfolio weights.

\subsubsection*{Combined estimators (UPSA–AO, AvgUPSA–AO)}

We use AO eigenvalues to filter all the covariance matrices used in UPSA: those defining ridge portfolios and those used within the cross-validation procedure. Note that AO filters the correlation matrices; the filtered covariance matrices are then reconstructed by rescaling with the original sample volatilities. 
We call the estimator obtained through this procedure \textbf{UPSA–AO}. We expect the time-variation of the optimal ridge weights to be appreciably smaller than with UPSA alone.

The \textbf{AvgUPSA–AO} estimator combines both enhancements: AO pre-filtering of correlation matrices and the averaging of UPSA-AO weights across past windows. 
In short, the AO step provides a robust noise reduction of correlation eigenvalues and some robustness to covariate shift prior to the UPSA stage, while the subsequent temporal averaging of UPSA weights further stabilizes the solution against fluctuations induced by grid discretization and sampling noise.

\section{Empirical Results}
\label{sec:sec3}
\ 
\subsection{Data}
We use the same dataset as \cite{NBERw32004}:  monthly returns of 153 characteristic-managed portfolios of US equities between 1970 and 2024 from  \cite{Jensen2023}.\footnote{Data available at: \url{https://jkpfactors.com/}} We consider five filtering schemes: UPSA, AvgUPSA, AO, UPSA-AO and AvgUPSA-AO. We define the OOS Sharpe ratio of a portfolio $\pi$, held for 6 months, with realized returns and risk:
\begin{equation}
\label{eq:sharpe_ratio}
    \mathrm{SR} = \frac{\boldsymbol{\pi}^\top \hat{\boldsymbol{\mu}}_{\mathrm{OOS}}}{\sqrt{\boldsymbol{\pi}^\top \hat{\boldsymbol{\Sigma}}_{\mathrm{OOS}} \,  \boldsymbol{\pi}}},
\end{equation}
where $\hat{\boldsymbol{\mu}}_{\text{OOS}}$ and $\hat{\boldsymbol{\Sigma}}_{\text{OOS}}$ are respectively the realized mean vector and the realized covariance matrix over the 6-month period after calibration of $\pi$. This definition of the Sharpe Ratio ensures a purely out-of-sample measurement of performance.

UPSA portfolios are computed using LOO CV; AvgUPSA adds an expanding-window mean of the optimized UPSA ridge weights. All of the Average Oracle eigenvalues use an exponentially weighted moving average of calibrated oracles with a half-life of 24 months. AO portfolios correspond to the efficient portfolios with AO-filtered correlation matrix (without UPSA). UPSA-AO and AvgUPSA-AO are obtained by pre-filtering all cross-sectional correlation matrices with AO eigenvalues, including the ones used during CV.

\subsection{Penalization grid}

\begin{figure}
    \centering
    \includegraphics[width=\textwidth]{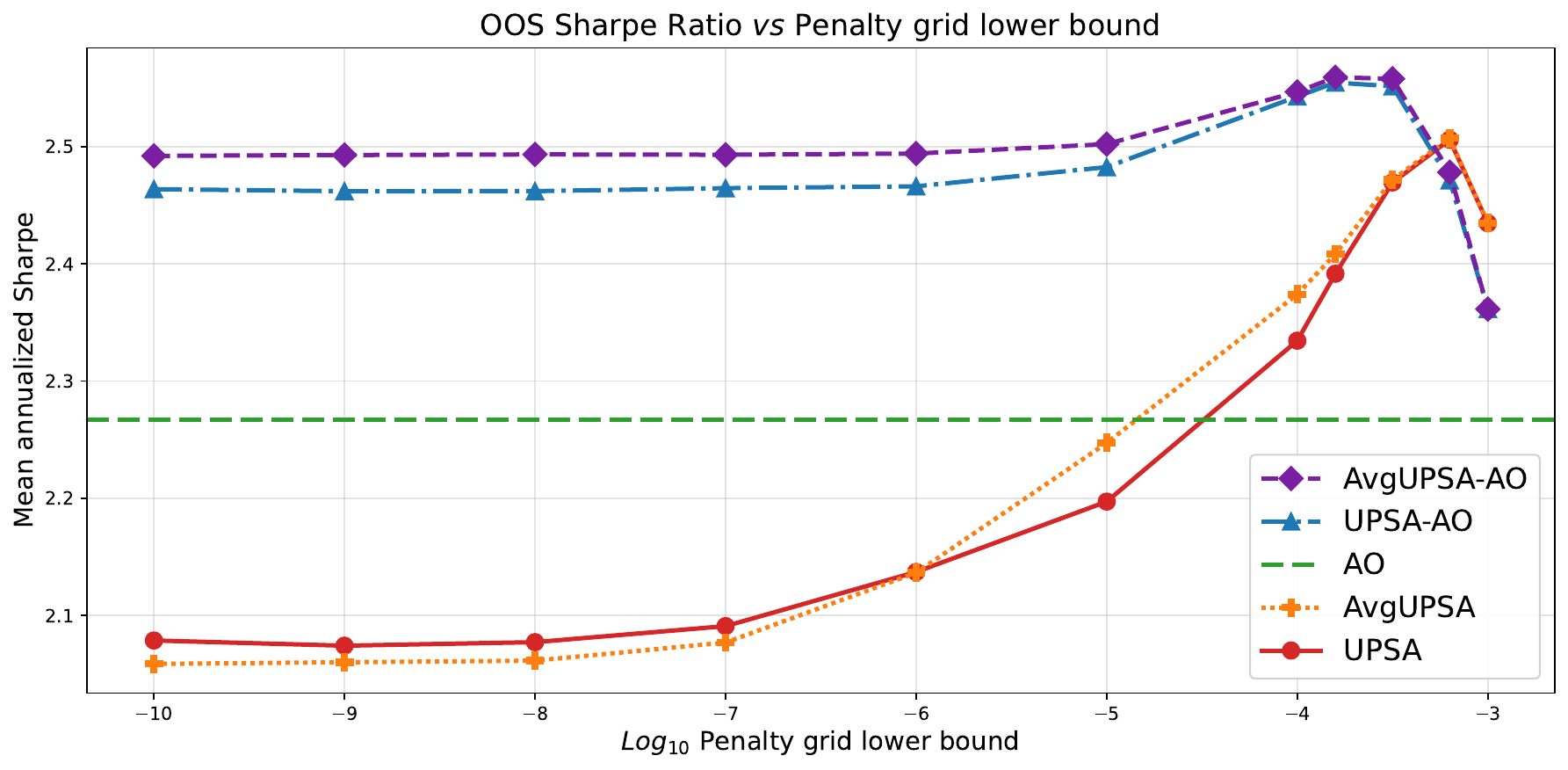}
    \caption{Mean Annualized Sharpe ratio against the penalty grid lower bound, for portfolios of the different proposed estimators on JKP US characteristic factors monthly returns, 153 assets, in-sample window size $T=120$M. Portfolios are rebalanced every month over the period 1985--2024. The realized OOS Sharpe Ratio is computed as the realized return over the 6 months divided by its realized volatility.}
    \label{fig:grid_sens}
\end{figure}

We first choose the same ridge penalty grid as \cite{NBERw32004}, which correspond to the left-most points in Figure \ref{fig:grid_sens}. All the AO-based filtering schemes over-perform vanilla UPSA and AvgUPSA in a statistically significant way. 

\cite{NBERw32004} recommend to use a grid that contains all the sample non-zero eigenvalues, which range from $10^{-7}$ to $10^{-1}$ in the considered data set. We found indeed that the grid upper bound  does not significantly affect the results provided that it is larger than the largest sample eigenvalue. However, the influence of the grid lower bound on UPSA and AvgUPSA performance is far from negligible. Figure \ref{fig:grid_sens} reports the OOS Sharpe ratio as a function of the grid lower bound. While the AO-based methods are only weakly sensitive to this choice, UPSA and AvgUPSA are much improved once the lower bound impose larger minimal penalization. However, setting the  grid lower bound to $10^{-4}$ is  unsatisfactory in principle, as this value roughly corresponds to the mean of the empirical eigenvalue distribution.  If the eigenvalues did not require any shrinkage, such a choice would therefore enforce an unnecessarily strong penalization, leading to suboptimal portfolios.

The sharp performance drop observed for AO-based estimators when the grid lower bound increases further is a direct consequence of this effect: because AO pre-filtering already compresses the smallest eigenvalues, imposing an excessively high minimal ridge penalty results in an over-shrunk covariance spectrum. 
In other words, the additional penalization acts on already filtered eigenvalues, effectively degrading precision rather than improving stability. Finally the large over-performance of UPSA-AO over UPSA is linked to the fact that the optimal ridge weights vary much less between two calibration windows (mean turnover of 0.35 vs 0.42, a highly significant difference) (see sec. \ref{sec:noise}). 

\subsection{Calibration window length}

We investigate the mean performance of the considered estimators on the 1985--2024 period with respect to the IS calibration window size $T$. We consider the same dataset as before and fix the ridge penalty grid to 20 logarithmically spaced points in $[10^{-8},10^{-1}]$. We run 13 experiments for $T$ varying between $3$ and $15$ years. In each run, Oracles are calibrated using folds $I_{\mathrm{cal}}$ and $ I_{\mathrm{test}}$ of respective sizes $T$ and $T_{\mathrm{OOS}} = 6$\ months. We can observe a consistent over-performance of combined estimators UPSA-AO and AvgUPSA-AO over UPSA.

\begin{figure}
    \begin{center} 
        \includegraphics[width=\textwidth]{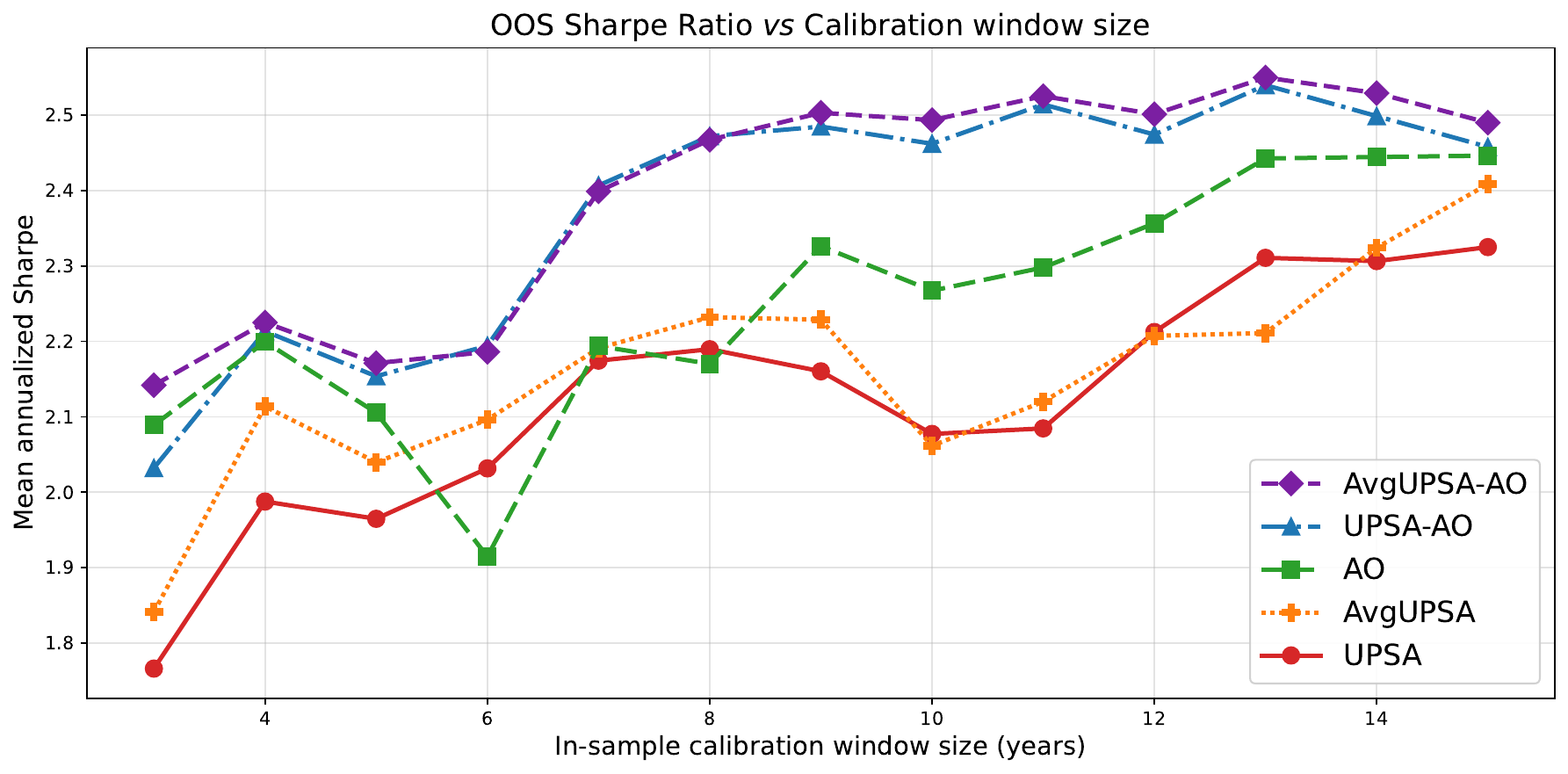} 
    \caption{Mean Annualized Sharpe ratio against the in-sample window size, using JKP US factors monthly returns, with $N=153$ assets and a penalty grid of 20 log-spaced points in [$10^{-8}$, $10^{-1}$].
        \\}
        \label{fig:complexity_loggrid} 
    \end{center}
\end{figure}

\begin{figure}
    \begin{center} 
        \includegraphics[width=0.95\textwidth]{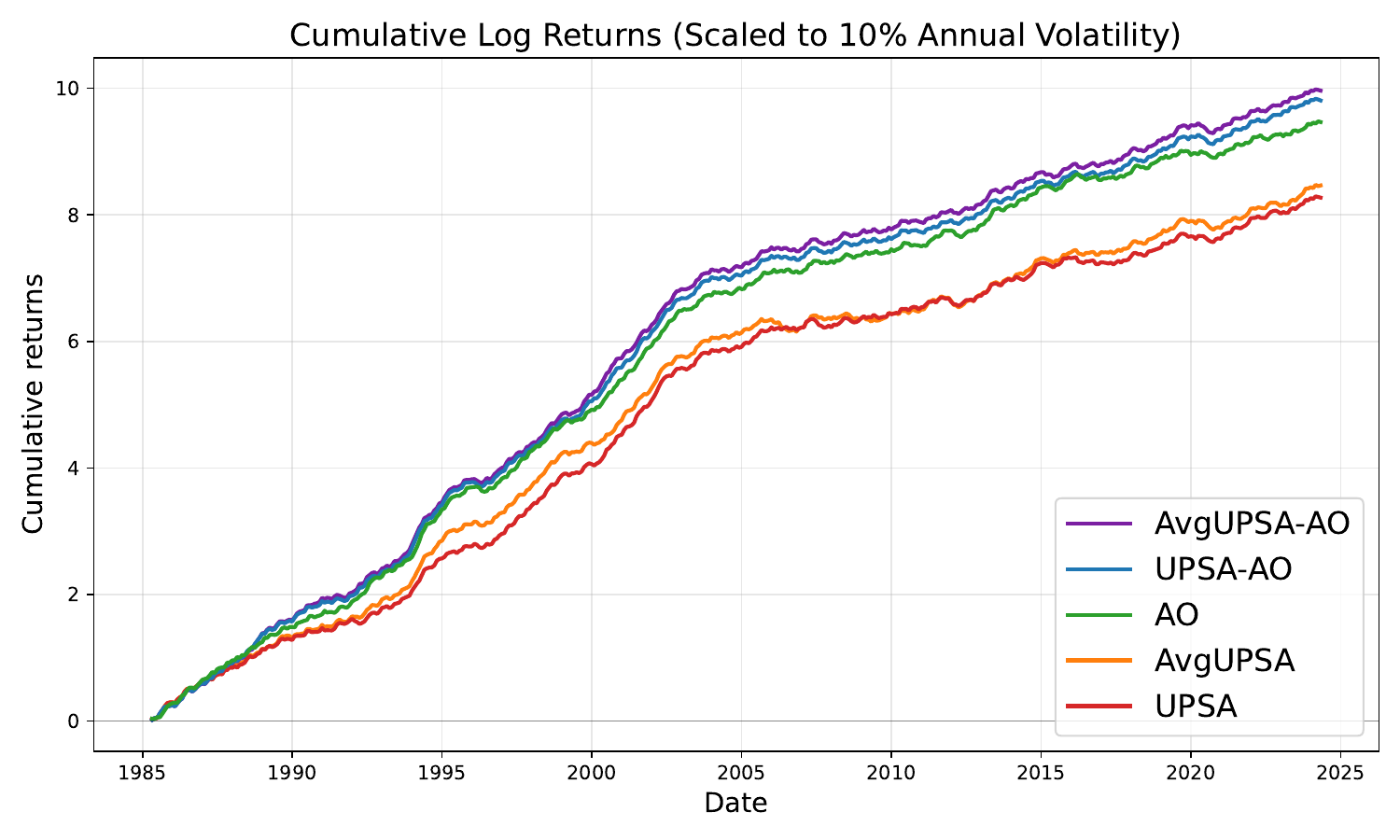} 
        \caption{Cumulative log-returns scaled by point-wise $6$M centered rolling standard deviations to achieve $10\%$ annual volatility, on JKP US factors' monthly returns with a fixed in-sample window size of T=120M,  and a penalty grid of 20 log-spaced points in [$10^{-8}$, $10^{-1}]$. The graph shows clear consistent historical over-performance of proposed AO-based estimators over UPSA.\\
        } 
        \label{fig:cumul_returns} 
    \end{center}
\end{figure}

The historical Sharpe ratio performance on the period 1985--2024 can also be visualized as the cumulative log-returns in Fig.\ \ref{fig:cumul_returns}. Portfolios are rebalanced every month, and their return is measured as the realized return over the following month. The resulting cumulative log-returns are rescaled using a centered 6-month rolling standard deviation, targeting an annualized volatility of 10\%. The in-sample calibration window is set to $T = 120$ months.

\subsection{Statistical analysis}

Table~\ref{table:tab1} reports the main statistics of the maximum–Sharpe portfolios obtained from the five estimators applied to monthly factor returns over the period 1985–2024, using an in-sample calibration window of $T=120$~months and a fixed grid of 20 logarithmically spaced penalty values between $10^{-8}$ and $10^{-1}$. 
All Sharpe ratios and associated statistical tests are computed from the out-of-sample Sharpe ratio defined in Eq.~\eqref{eq:sharpe_ratio}. 
The table displays, in order, the mean annualized Sharpe ratio, average diversification (inverse of the sum of squared portfolio weights), mean turnover, gross leverage, maximum monthly drawdown, and the outcomes of the statistical tests.

Using  one-sided Wilcoxon signed-rank tests for performance comparisons, we find that the proposed AvgUPSA–AO estimator achieves the highest out-of-sample Sharpe ratio, outperforming all others, including the second best. UPSA–AO, with a $p$–value of $0.012$. 
It is also superior to the other estimators in all criteria except for the maximum drawdown. 
In contrast, AvgUPSA is found to perform slightly worse than UPSA under this configuration, though this difference is not statistically significant ($p$–value  equal to 0.38). 
However, for alternative penalty grids with higher lower bounds, AvgUPSA  outperforms UPSA. In fact AvgUPSA is consistently better than UPSA when considering monthly rebalancing of the portfolios and for every penalty grid, as can be observed in Figure \ref{fig:cumul_returns}.

To ensure robustness in the statistical assessment, we further apply the Model Confidence Set (MCS) procedure  \citep{MCS}, which identifies the subset of estimators that are statistically indistinguishable from the best-performing one, using the negative out-of-sample Sharpe ratio as loss function.
The MCS results confirm that both {UPSA–AO} and {AvgUPSA–AO} belong to the superior set of models (test size of $0.05$). , confirming their advantage over the other estimators.

In summary, the AvgUPSA–AO estimator demonstrates a statistically significant and robust improvement over all alternatives. 
Its performance gains stem from the combination of the AO pre-filtering, which stabilizes correlation estimation, and the averaging of optimized weights. Quite notably, it is better than either UPSA or AO of their own.

\begin{table}
  \centering
  \scriptsize
  \resizebox{\textwidth}{!}{%
    \begin{filecontents*}{csv/table.csv}
,UPSA,AvgUPSA,AO,UPSA-AO,AvgUPSA-AO
Mean Annualized Sharpe,2.077,2.061,2.267,\textbf{2.462},\textbf{2.493}
Mean Diversification,6.7,3.8,6.1,16.4,\textbf{17.7}
Mean Turnover,1.71,1.27,0.67,0.31,\textbf{0.25}
Mean Gross Leverage,5.38,6.01,4.18,2.50,\textbf{2.32}
Max Drawdown,-2.66,-4.64,\textbf{-2.36},-2.65,-2.73
Wilcoxon p-value (Est - UPSA),/,3.8e-01,5.31e-03,2.1e-10,3.1e-11
Wilcoxon p-value (AvgUPSA-AO - Est),3.1e-11,2.8e-10,8.1e-05,1.2e-02,/
MCS included (size 0.05),No,No,No,Yes,Yes
\end{filecontents*}
\pgfplotstabletypeset[
      every first column/.append style={column name={Key statistics / Estimators}, column type=l|},
      every row/.style={before row={\rule{0pt}{2.5ex}}},
      col sep=comma,
      header=true,
      every head row/.style={before row=\toprule,after row=\midrule},
      every last row/.style={after row=\bottomrule},
      columns/UPSA/.style={string type,column name={UPSA}},
      columns/AvgUPSA/.style={string type,column name={AvgUPSA}},
      columns/AO/.style={string type,column name={AO}},
      columns/UPSA-AO/.style={string type,column name={UPSA-AO}},
      columns/AvgUPSA-AO/.style={string type,column name={AvgUPSA-AO}},
    ]{csv/table.csv}
  }
  \caption{Key statistics of efficient Max Sharpe portfolios estimated on monthly factor returns over the 1985--2024 period. 
  Reported quantities are, from top to bottom: mean annualized Sharpe ratio as computed in Eq. \ref{eq:sharpe_ratio}, average diversification (inverse sum of squared weights), mean portfolio turnover, gross leverage, and maximum monthly drawdown. 
  The last three rows show one-sided Wilcoxon signed-rank tests comparing Sharpe ratios of estimators {\em vs.} UPSA, AvgUPSA-AO {\em vs.} estimators and the Model Confidence Set (MCS) inclusion at 5\% test size.\\
  Portfolios are optimized with an in-sample window size of $T=120$ months.
  The ridge penalty grid contains 20 logarithmically spaced values in $[10^{-8}, 10^{-1}]$.}
  \label{table:tab1}
\end{table}

\section{Conclusion}

Despite its undeniable filtering abilities, UPSA suffers from too much noise and a strong dependence on the choice of penalty grid. Filtering noise further increases the abilities of UPSA. 
First, using long time averages of past optimal ridge weights improves the performance of UPSA for well chosen grids.
Second, pre-filtering covariance matrices with Average Oracle provides a strong baseline, effectively stabilizing covariance estimates and accounting partly for covariate shift before the UPSA refinement.  
Finally, compounding averaging and AO filtering yields the novel estimator AvgUPSA-AO that inherits the regularization benefits of both approaches and outperforms statistically significantly all other tested methods across most settings. 

From a broader perspective, the UPSA part of AvgUPSA-AO acts as a first-order improvement over AO, a zeroth-order nonlinear filtering method. This opens the question of using the UPSA technique to optimize  the optimal oracle values directly, as well as using adaptive grid bounds. Other future directions include exploring new parametrized families of covariance filtering, such as the James–Stein shrinkage for eigenvectors by \cite{GolbergJSE}.

\section*{Acknowledgments}

P.R. acknowledges funding from ANRT, under the CIFRE contract nr 2025/0279.

This publication used HPC resources from the “Mésocentre” computing center of CentraleSupélec and École Normale
Supérieure Paris-Saclay supported by CNRS and Région Île-de-France.

\bibliographystyle{plainnat}  
\bibliography{references}  

\end{document}